\documentclass[prb,floatfix,twocolumn,showpacs,superscriptaddress]{revtex4}
\usepackage{amsmath}
\usepackage{graphicx}
\usepackage{dcolumn}
\usepackage{hyperref}
\newcommand{\smallwidth}{0.7\columnwidth}
\newcommand{\figwidth}{0.95\columnwidth}

\newcommand{\PP}{\mathcal{P}}
\newcommand{\PJ}{\mathcal{P}_{\mathcal{J}}}
\newcommand{\DID}{d_{x^{2}-y^{2}}+id_{xy}}
\newcommand{\var}{\mathop{\rm var}}
\newcommand{\tj}{$t{-}J$ }
\newcommand{\QQ}{\mathbf{Q}_N=(\pi,-\pi/\sqrt{3})}

\begin{document}
\title{Magnetism and superconductivity of strongly correlated
  electrons\\ on the triangular lattice}
\author{C\'edric Weber}
\author{Andreas Laeuchli}
\affiliation{Institut Romand de Recherche Num\'erique en Physique
  des Mat\'eriaux (IRRMA), CH-1015 Lausanne, Switzerland}
\author{Fr\'ed\'eric Mila}
\affiliation{Institute of Theoretical Physics, EPFL, CH-1015
  Lausanne, Switzerland}
\author{Thierry Giamarchi}
\affiliation{DPMC, University of Geneva, Quai Ernest Ansermet 24,
  CH-1211 Geneva, Switzerland}
\date{\today}

\begin{abstract}
  We investigate the phase diagram of the \tj Model on a triangular
  lattice using a Variational Monte-Carlo approach.
  We use an extended set of Gutzwiller projected fermionic trial wave-functions
  allowing for simultaneous magnetic and superconducting order parameters.
  We obtain energies at zero doping for the spin-$1/2$ Heisenberg
  model in very good agreement with the best estimates.
  Upon electron doping (with a hopping integral $t<0$)
  this phase is surprisingly stable variationally
  up to $n\approx 1.4$, while the $d_{x^{2}-y^{2}}+i d_{xy}$
  order parameter is rather weak and disappears
  at $n\approx 1.1$. For hole doping however the coplanar
  magnetic state is almost immediately destroyed and $d_{x^{2}-y^{2}}+i d_{xy}$
  superconductivity survives down to
  $n\approx 0.8$. For lower $n$, between $0.2$ and $0.8$, we find saturated ferromagnetism.
  Moreover, there is evidence for a narrow spin density wave phase around $n\approx 0.8$.
  Commensurate flux phases were also considered, but these turned out {\em not} to be
  competitive at finite doping.
\end{abstract}
\maketitle

\section{Introduction}
\label{intro}

The discovery of high-$T_c$ superconductivity, and the
observation\cite{anderson_hgtc_hubbard} that strong correlations are
important in connection with these compounds has led to a tremendous
interest in understanding strongly correlated electron physics. In
particular the two simplest models for strongly correlated
electrons, namely the Hubbard and \tj models, have been the subject
of intensive studies.

For example, one question of crucial interest is the interplay
between superconductivity and antiferromagnetism close to the
insulating phase in the \tj model. The ground state of this model
on the square lattice is known to be antiferromagnetic at
half-filling and one of the important questions is what happens
upon doping. All the approaches to these strong coupling problems
involve approximations, and it is sometimes difficult to
distinguish the artefact due to approximations from the true
features of the model. However, for the case of the square
lattice, both the variational Monte-Carlo method
(VMC)\cite{yokoyama_mcv_supra,gros_mcv_supra} and mean-field
theories\cite{kotliar_fluxphases} have found a d-wave
superconducting phase in the the \tj model. A wavefunction
combining antiferromagnetism and superconductivity was proposed
for the Hubbard and \tj
models\cite{giamarchi_coexistence_comm,giamarchi_tjhub}, allowing
to reconcile the variational results between these two models.
This wavefunction allowed for an excellent variational energy and
order parameter and a range of coexistence between
superconductivity and anti-ferromagnetism was found. Further
investigations of this class of wavefunctions has been very
fruitful for the square lattice. This allowed to successfully
compare to some of the experimental features with the high-$T_c$
cuprates\cite{paramekanti_vmc,anderson_RVB_AtoZ}, even if of
course many questions remain regarding the nature of the true
ground state of the system.

The resonating valence bond (RVB) scenario proposed by Anderson
\cite{anderson_hgtc_hubbard} was argued to be even more relevant
in the geometry of the triangular lattice. At half-filling, the
lattice is a frustrated magnet: the competition between the
exchange integrals leads to unsatisfied bonds. The original
expectation is that quantum fluctuations might lead to a
spin-liquid behavior. However, at half-filling, it appears by now
that the spin-$1/2$ triangular lattice has a three sublattice
coplanar magnetic order
\cite{huse_triangular,bernu_exact_diag_triangular_short,capriotti_spin_heisenberg_triangular}.
Quantum fluctuations are nevertheless strong, and the sublattice
magnetization is strongly reduced due to these fluctuations. It is
therefore expected that the magnetism is fragile and quickly
destroyed by doping and that a strong RVB instability is present.
Indeed, RVB mean field theories
\cite{baskaran_mf_triangular,kumar_mf_triangular,wang_mf_triangular}
were used for the \tj model, and $d_{x^{2}-y^{2}}+i d_{xy}$
pairing was found over a significant range of doping. The same
approach and questions that arise in the framework of the \tj
model on the square lattice are thus very relevant in the present
frustrated lattice. The success of the variational approach for
the square lattice suggests to investigate the same class of
variational wavefunctions for the triangular one.

Besides its own theoretical interest, another motivation for
understanding the physics of electrons on a triangular lattice is
provided by the recent discovery of superconductivity at low
temperature in the CoO$_{2}$ layered compounds
\cite{takada_cobaltites_discovery}
(Na$_{\delta}$CoO$_{2}$.yH$_{2}$O). In these systems,
superconductivity is observed in a range of electron doping
$\delta$ between $25\%$ and $33\%$
\cite{schaak_discovery_Tc_cobaltites}.
Na$_{\delta}$CoO$_{2}$.yH$_{2}$O consists of two dimensional
CoO$_{2}$ layers separated by thick insulating layers of Na$^{+}$
ions and H$_{2}$O molecules. It is a triangular net of edge
sharing oxygen octahedra;\ Co ions are at the center of the
octahedra forming a 2D triangular lattice. Takada et al.
\cite{takada_cobaltites_discovery} speculated that this system
might be viewed as a doped spin-$1/2$ Mott insulator. Based on LDA
calculations \cite{singh_lda_cobaltites}, a simplified single band
\tj picture with negative $t$ and electron doping was put forward
\cite{baskaran_mf_triangular,kumar_mf_triangular,wang_mf_triangular}.
Such systems might thus be the long-sought low-temperature
resonating valence bond superconductor, on a lattice which was at
the basis of Anderson's original ideas on a possible quantum spin
liquid state
\cite{AndersonRVBTriangular,FazekasAndersonRVBTriangular}.

We propose in this paper to study the \tj model within the
framework of the Variational Monte-Carlo (VMC) method, which
provides a variational upper bounds for the ground state energy.
In contrast to mean-field theory, it has the advantage of exactly
treating the no double-occupancy constraint. VMC using simple RVB
wave-functions has been used for the triangular lattice
\cite{watanabe_VMC_TJ_triangular} and it was found that $\DID$
superconductivity is stable over a large range of doping. However,
in the previous study the fact that the \tj is magnetically
ordered at half-filling was not taken into account. We expect that
the frustration in the triangular lattice may lead to a richer
phase diagram and to many different instabilities. We thus propose
here to study extended wave-functions containing at the same time
magnetism, flux phase and RVB instabilities, in a similar spirit
as for the square
lattice\cite{giamarchi_coexistence_comm,giamarchi_tjhub}, in order
to study in detail the interplay between frustrated magnetism and
superconductivity. Given the non-collinear nature of the magnetic
order parameter compared to the case of the square lattice, the
task is however much more complicated. We thus present in this
work a general mean-field Hamiltonian which takes into account the
interplay between magnetic, RVB and flux-phase instabilities. The
resulting variational wave-function is sampled with an extended
VMC, which uses {\em Pfaffian} updates rather than the usual
determinant updates. We show that the interplay between the
different instabilities leads to a faithful representation of the
ground-state at half-filling, and we also find good variational
energies upon doping. To benchmark our wave-function, we carry out
exact diagonalizations on a small $12$ sites cluster and compare
the variational energies and the exact ones. Finally, a
commensurate spin density wave is considered, and is shown to be
relevant for the case of hole doping.

The outline of the paper is as follows: in Sec.~\ref{sec:model} we
present the model and the numerical technique. In
Sec.~\ref{sec:results} we show the variational results both for the
case of hole and electron doping. Finally Sec.~\ref{sec:conclusion}
is devoted to the summary and conclusions.

\section{Model and method} \label{sec:model}

We study the \tj model on the triangular lattice defined by the
Hamiltonian:
\begin{multline}\label{Htj}
  H_{t-J} = -t\sum\limits_{\langle i,j \rangle,\sigma }{
    \left( c_{i,\sigma }^{\dagger}c_{j,\sigma} + h.c. \right)}  \\
  +J\sum\limits_{\langle i,j \rangle}{  \left( \mathbf{S}_{i}\cdot
      \mathbf{S}_{j}-\frac{{1}}{4}n_{i}n_{j} \right)}
\end{multline}
The model describes electrons hopping with an amplitude $t$, and
interacting with an antiferromagnetic exchange term $J$ between
nearest neighbor sites (denoted $\langle i,j \rangle$) of a
triangular lattice. $\mathbf{S}_{i}$ denotes the spin at site $i$,
$\mathbf{S}_{i}=\frac{1}{2}c_{i,\alpha}^{\dagger}\vec{\sigma}_{\alpha,\beta}
c_{i,\beta}$ and $\vec{\sigma}$ is the vector of Pauli matrices.
$H_{t-J}$ is restricted to the subspace where there are no doubly
occupied sites. In order to simplify the connection to the
Cobaltates we set $t=-1$ in the following and present the results
as a function of the electron density $n \in [0,2]$, half-filling
corresponding to $n=1$. $n > 1$ corresponds to a \tj model at
$\tilde{n}=2-n$ for $t=1$, by virtue of a particle-hole
transformation. In the first part of this section, we emphasize on
the method to construct a variational wave-function containing
both superconductivity and non-collinear magnetism. The
wave-function allows to consider $3$-sublattice magnetism,
however, since the latter wavefunction is restricted to a $3$ site
supercell, we briefly comment on a second simpler variational
wave-function type, which allows to describe commensurate spin
order. In the second part of the section, we define the relevant
instabilities and the corresponding order parameters.

\subsection{Variational wave-function}

\begin{figure}
  \begin{center}
    \includegraphics[width=\smallwidth]{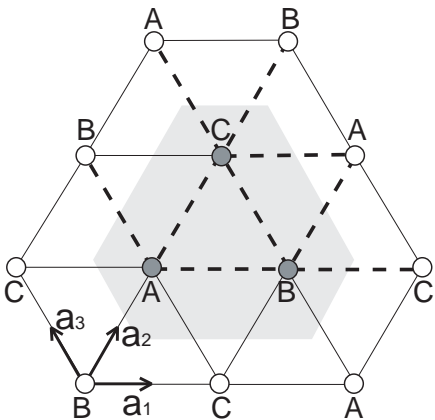}
    \caption{$3$-site supercell of the triangular lattice.
     The onsite magnetic variational parameters can vary
     independently on each of the site A,B and C of the supercell.
     The BCS pairing as well as the flux vary independently on
     each of the different dashed bonds.}
    \label{fig:lattice}
  \end{center}
\end{figure}

In order to study this model we use a variational wavefunction built
out of the ground state of the following mean-field like
Hamiltonian:
\begin{multline}
\label{eq:HMF}
  H_{MF}=\sum\limits_{\langle i,j \rangle,\sigma}
  \left( -t e^{i\theta_{i,j}^{\sigma}} c_{i\sigma }^{\dagger} c_{j\sigma} + h.c.\right) \\
  +\sum\limits_{\langle i,j \rangle,\sigma,\sigma^{\prime }} \left(
    \left\{\Delta_{\sigma ,\sigma^{\prime}}\right\}_{i,j}c_{i\sigma}^{\dagger}
    c_{j\sigma^{\prime}}^{\dagger}
    + h.c.\right)\\
  +\sum\limits_{i} \mathbf{h}_{i}\cdot\mathbf{S}_{i}
  -\mu\sum\limits_{i,\sigma } n_{i,\sigma }
\end{multline}
$H_{MF}$ contains at the same time BCS pairing
($\mathbf{\Delta}_{i,j}=\left\{\Delta_{\sigma,\sigma^{\prime}}\right\}_{i,j}$),
an arbitrary external magnetic field ($\mathbf{h}_{i}$), and
arbitrary hopping phases ($\theta_{i,j}^{\sigma}$), possibly spin
dependent. These variational parameters are unrestricted on the
$A,B,C$ sites and the corresponding bonds of a $3$-site supercell,
as shown in Fig.~\ref{fig:lattice}.
We allow both singlet ($\mathbf{\Delta}_{i,j}^{(S=0)}$) and general
triplet ($\mathbf{\Delta}_{i,j}^{(S=1)}$) pairing symmetries to be
present. They correspond to choosing:
\begin{equation}
\begin{split}
  \mathbf{\Delta}_{i,j}^{(S=0)} &= \left({\begin{array}{*{20}c}
        0        & \psi_{i,j}        \\
        { - \psi_{i,j} } &   0         \\\end{array}}\right) \\
  \mathbf{\Delta}_{i,j}^{(S=1)} &= \left({\begin{array}{*{20}c}
        \psi^2_{i,j}        & \psi^1_{i,j}        \\
        { \psi^1_{i,j} } &   \psi^3_{i,j}         \\\end{array}}\right)
\end{split}
\end{equation}

Since $H_{MF}$ is quadratic in fermion operators it can be solved by
a Bogoliubov transformation. In the most general case considered
here, this gives rise to a $12\times12$ eigenvalue problem, which we
solve numerically. We then find the ground state of $H_{MF}$
\begin{multline}
\left| {\psi_{MF} } \right \rangle = \exp \left\{
{\sum\limits_{i,j,\sigma _i
        ,\sigma _j } {a_{(i,j,\sigma _i ,\sigma _j )} c_{i\sigma _i }^\dagger
        c_{j\sigma _j }^\dagger } } \right\} \left| 0 \right \rangle
\end{multline}
Here $a_{(i,j,\sigma _i ,\sigma _j)}$ are numerical coefficients.
Note that $\left| \psi_{MF} \right\rangle$ has neither a fixed
number of particles due to the presence of pairing, nor a fixed
total $S^z$ due to the non-collinear magnetic order. Thus in order
to use it for the VMC study we apply to it the following
projectors: $\PP_N$ which projects the wave-function on a state
with fixed number of electrons and $\PP_{S^z}$ which projects the
wave-function on the sector with total $S^z=0$. Finally we discard
all configurations with doubly occupied sites by applying the complete
Gutzwiller projector $\PP_\mathcal{G}$. The wavefunction we use as an input
to our variational study is thus:
\begin{multline}\label{eq:startfunc}
  \left| \psi_{\var} \right\rangle = \PP_{\mathcal{G}} \PP_{S^z}  \PP_{N}
  \left| {\psi _{MF} } \right\rangle \\
  =\PP_{\mathcal{G}} \PP_{S^z} \left\{ {\sum\limits_{i,j,\sigma _i ,\sigma
        _j } {a_{(i,j,\sigma _i,\sigma _j )} c_{i\sigma _i }^\dagger
        c_{j\sigma_j }^\dagger }} \right\}^{N/2}  \left| 0 \right \rangle
\end{multline}
Although the wavefunction (\ref{eq:startfunc}) looks formidable, it
can be reduced to a form suitable for VMC calculations. Using
\begin{equation}
  \left\langle \alpha \right\vert
  =\left\langle 0\right\vert c_{k_{1},\sigma_{1}}...c_{k_{N},\sigma_{N}},
\end{equation}
we find that
\begin{equation}
\begin{split}
  \left\langle {\alpha }\mathrel{\left | {\vphantom {\alpha {\psi _{MF} }}}
      \right. \kern-\nulldelimiterspace} {{\psi _{\var} }}\right\rangle
  &= P_{f}\left( \mathbf{Q}\right) \\
  Q_{i,j} &= a_{(k_{i},k_{j},\sigma _{i},\sigma_{j})}
  -a_{(k_{j},k_{i},\sigma _{j},\sigma _{i})}
\end{split}
\end{equation}
where $P_{f}(\mathbf{Q})$ denotes the \emph{Pfaffian} of the
matrix $\mathbf{Q}$. Using this last relation, the function
(\ref{eq:startfunc}) can now be evaluated numerically using a
Monte Carlo procedure with Pfaffian updates, as introduced in
Ref.~\onlinecite{bouchaud_supra_determinant}. In the particular
case where
$a_{k,l,\uparrow,\uparrow}=a_{k,l,\downarrow,\downarrow}=0$ and
at $S^z=0$ (this happens if the BCS pairing is of singlet type
{\em and} the magnetic order is collinear),
the Pfaffian reduces to a simple
determinant, and the methods becomes equivalent to the standard
Variational Monte-Carlo\cite{ceperley_mcv_fermions} technique.

The above mean field Hamiltonian and wavefunction contain the main
physical ingredients and broken symmetries we want to implement in
the wavefunction. In order to further improve the energy and allow
for out of plane fluctuations of the magnetic order we also add a
nearest-neighbor spin-dependent Jastrow \cite{jastrow_fonctions}
term to the wave-function:
\begin{equation}\label{eq:jastrow}
  \PJ=\exp{\left( \alpha\sum\limits_{ \langle i,j \rangle }{S_i^z S_j^z }\right)},
\end{equation}
where $\alpha$ is an additional variational parameter. Our final
wavefunction is thus:
\begin{equation} \label{eq:wavefinal}
  \left|\psi_{\var} \right\rangle = \PJ  \PP_{S^z} \PP_{N}
  \PP_{\mathcal{G}} \left| {\psi _{MF} } \right\rangle
\end{equation}
When $\alpha<0$ the Jastrow factor favors all configurations which
belong to the ground state manifold of a classical Ising
antiferromagnet on the triangular lattice. Such a manifold is
exponentially large \cite{wannier_triangular_magnet_Ising}, and
this Jastrow factor thus provides a complementary source of spin
fluctuations.

In what follows we use the wavefunction (\ref{eq:wavefinal})
directly for the VMC, but we also examine improved wavefunctions
with respect to (\ref{eq:wavefinal}) that can be obtained by
applying one or more Lanczos steps
\cite{heeb_lanczos_step_epl,heeb_thesis,becca_thesis}:
\begin{equation}
  \left| {1Ls} \right\rangle = \small(
  {1 + \lambda H_{t - J} } \small) \left| \psi_{\var} \right\rangle
\end{equation}
with optimized\cite{sorella_1Lanczos_step} $\lambda$. Since the
calculation of Lanczos step wave functions beyond the first step
is very time consuming, most of the results we will present here
were obtained using a single Lanczos step.

In the following, to clearly indicate which wavefunction we use,
we will denote them in the following way: $MF\ /\ J\ /\ n Ls$,
where MF denotes the fields present in the mean-field like
Hamiltonian $H_{MF}$, $J$ is present if we use the Jastrow factor,
$n\ Ls$ denotes the presence and the number of Lanczos steps
applied on top of the bare wave function.

As usual with the VMC procedure, these general wave functions are
now used to minimize the expectation value of the total energy
$\langle H_{t{-}J}\rangle$ by changing the variational parameters.
We used a correlated measurement technique
\cite{umrigar_mcv_minimis,giamarchi_coexistence_comm,giamarchi_tjhub}
combined with parallel processing to smoothen the energy landscape
and use a steepest-descent type routine to locate the minimum of
energy. We then define the \emph{condensation energy} $e_c$ of the
optimal wave function as
\begin{equation}
  e_{c}=e_{\text{var.}}-e_{\text{Gutzwiller}},
\end{equation}
where $e_{\text{Gutzwiller}}$ is the energy of the Gutzwiller wave
function, i.e. the fully projected Fermi sea at zero magnetization.
In some cases we had to keep a small BCS pairing field to avoid
numerical instabilities. Let us note that the linear size of the
$\mathbf{Q}$ matrix is two times larger than in the simpler case of
determinantal update VMC. Therefore our largest $108$ sites cluster
with Pfaffian updates corresponds roughly to a $200$ sites cluster
using standard updates.

\subsection{Commensurate order}

 Since the mean-field Hamiltonian (\ref{eq:HMF}) is restricted to
 a $3$ site supercell, we investigated also a second class of mean-field
 Hamiltonians based on collinear commensurate structures, which are not contained
 in the previous Hamiltonian.
 For this type of phase, we used a simpler mean-field ansatz along the lines
 of Ref.~\onlinecite{giamarchi_tjhub}. The mean-field
 Hamiltonian written in k-space is
\begin{multline}
\label{eq:HSDW} H_{SDW}  = \sum\limits_{k,\sigma}
  \left(  \left( \epsilon_k-\mu \right) c_{k\sigma }^{\dagger}
  c_{k\sigma}+ f(Q,\sigma)c_{k+Q\sigma }^{\dagger} c_{k\sigma}
  \right) \\
+ \sum\limits_{k} \left( \Delta_k
c^{\dagger}_{k\uparrow}c^{\dagger}_{-k\downarrow} +h.c.\right),
\end{multline}
where k does run over the Brillouin zone of the original
triangular lattice, $\epsilon_k$ is the dispersion of the free
electron Hamiltonian, and $\Delta_k$ is the Fourier transform of
$\Delta_{i,j}$. Depending on $f(Q,\sigma)$, the ground state of
the Hamiltonian is a commensurate charge density wave
($f(Q,\sigma)=f(Q)$) or a spin density wave ($f(Q,\sigma)=\sigma
f(Q)$).

At half-filling, we considered also several commensurate flux
phases with $2\pi \times \frac{q}{p}$ flux per plaquette,
using the Landau gauge
\footnote{$H_{MF}$ is gauge invariant,
  however the gauge plays a non trivial role in the projected
  wave-function. Indeed, the kinetic energy of the projected
  wave-function depends on the chosen gauge. One cannot exclude
  that another choice of the gauge could lead to a better
  wave-function.
} with $p\in\{2,\dots,10\}$ and $q<p$. The one with the lowest
energy was found to be the $q=1,p=4$, as predicted theoretically
\cite{rokhsar_fluxphases_triangular}, giving an energy close to
the simple $\DID$ wave-function. Upon doping however the energy of
such commensurate flux phases are rapidly much worse than the
energies of our best wave functions. The main reason for this poor
performance upon doping is the rather bad kinetic energy of these
wave functions.

\subsection{Characterization of the encountered instabilities}

By minimizing all the variational parameters of the mean-field
Hamiltonians (\ref{eq:HMF}) and (\ref{eq:HSDW}) on a $12$ and a
$48$ site lattice, we find that the relevant instabilities present
at the mean field level consist of:
\begin{figure}
  \begin{center}
    \includegraphics[width=\smallwidth]{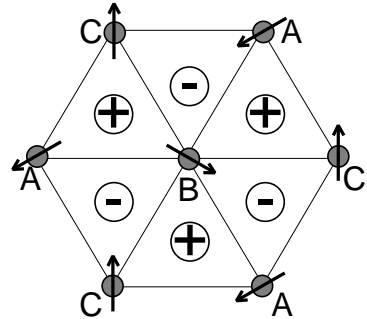}
    \caption{
      The variational parameters $\mathbf{h}_i$
      for the coplanar $120^\circ$ antiferromagnetic order.
      The spins lie in the $x-y$ plane.
      The $z$-component of the vector chirality ($\pm1$) on each
      triangular plaquette forms a staggered pattern.
      \label{fig:flux}}
  \end{center}
\end{figure}
\begin{itemize}
\item
  a $120^\circ$ coplanar antiferromagnetic order ($AF$),
  represented in Fig.~\ref{fig:flux}.
\item
  a concomitant staggered spin flux phase instability ($SFL$) with:
  \begin{align*}
    &\theta_{i,i+a_{1},\sigma}= \phantom{-}\theta \sigma \\
    &\theta_{i,i+a_{2},\sigma}=          - \theta \sigma \\
    &\theta_{i,i+a_{3},\sigma}= \phantom{-}\theta \sigma.
  \end{align*}
  These bond phase factors correspond to a spin current
  in the $z$ direction which is staggered on elementary triangles
  of the triangular lattice. This instability follows rather
  naturally, since the $120^\circ$ antiferromagnetic state itself
  already displays the same staggered spin currents
  $\left(\mathbf{S}_{i}{\times}\mathbf{S}_{j}\right)^{z}$ on the
  nearest neighbor bonds (see Fig.~\ref{fig:flux}). The effect
  of this instability was rather small and visible only at
  half-filling.
\item
  a translationally invariant superconducting phase with
  $d_{x^{2}-y^{2}}+id_{xy}$ singlet pairing symmetry ($d^+$),
  as well as the $d_{x^{2}-y^{2}}-id_{xy}$ ($d^{-}$). We
  have also looked extensively for triplet pairing for both
  electron and hole dopings and low $J/|t| \leq 0.4 $ on a 48
  site cluster, but with no success. The minimum energy was
  always found for singlet pairing symmetry.
\item
  a ferromagnetic state with partial or full polarization ($F$),
\item
  a commensurate collinear spin density wave~\cite{honerkamp_triangular}
 ($SDW$) instability with wavevector $\QQ$.
\end{itemize}

\subsection{Order parameters}

 In order to characterize the phases described by the
optimal wave functions after projection, we have calculated the
following observables:
\begin{itemize}
\item
  the sublattice magnetization of the $120^\circ$ coplanar antiferromagnetic
  order:
  \begin{equation}
    M_{AF}=
    \frac{1}{N}\sum\limits_{i}\left\| \frac{{\left\langle {\psi _{\var}|\mathbf{S}%
              _{i}|\psi _{\var}}\right\rangle }}{{\left\langle {{\psi _{\var}}}
            \mathrel{\left | {\vphantom {{\psi_{\var} } {\psi _{\var}}}} \right.
              \kern-\nulldelimiterspace}{{\psi _{\var}}}\right\rangle
        }}\right\|
  \end{equation}
  We have checked that the projected magnetization has the correct $120^\circ$ symmetry.
  To simplify the calculations, this expectation value has been sampled using wavefunctions
  without the projector $\PP_{S^z}$. We have checked that this gives the same result
  as the correlation function
  $\lim_{r \rightarrow \infty}\sqrt{\mathbf{S}_i\cdot\mathbf{S}_{i+r}}$ calculated with
  the projector $\PP_{S^z}$.
\item
  the $z$ component of the vector chirality (spin twist) on nearest neighbor bonds:
  \begin{equation} \chi=\frac{1}{3N}\sum\limits_{i,\alpha} \left|
      \frac{{\left\langle{\psi _{\var}|\left(\mathbf{S}_{i}{\times } \mathbf{S}_{i{+a }_{\alpha
                  }}\right) ^{z}|\psi _{\var}}\right\rangle
        }}{{\left\langle {{\psi _{\var}}}\mathrel{\left|
                {\vphantom {{\psi _{\var}} {\psi_{\var}}}} \right. \kern-\nulldelimiterspace}
            {{\psi_{\var} }}\right\rangle }}\right|
  \end{equation}
  We have checked that the measured vector chirality has the symmetry of the staggered
  currents derived from the 120$^\circ$ coplanar structure (c.f. Fig.~\ref{fig:flux}).
\item
  the amplitude of the absolute value of the collinear magnetization in the spin density wave
  wave-function :
 \begin{equation}
  M_{SDW}= \frac{1}{N}\sum\limits_{i} \left| \frac{ \langle \psi_{\var}  \left| S^{z}_{i} \right| \psi_{\var} \rangle }
  {\langle \psi_{\var} | \psi_{\var}\rangle}\right|
 \end{equation}

\item
  the amplitude of the singlet superconducting order parameter:
  \begin{equation}
    \Delta=\sqrt{\frac{1}{4N}\left|\mathop {\lim }
        \limits_{r\rightarrow \infty }\sum\limits_{i}\frac{{\left\langle {\psi _{{
                    \mathop{\rm var}}}|\Delta _{i,\alpha }^{\dagger}\Delta _{i+r,\beta }|\psi _{{
                    \mathop{\rm var}}}}\right\rangle }}{{\left\langle {{\psi _{{\mathop{\rm var}}
                  }}}\mathrel{\left | {\vphantom {{\psi _{{\mathop{\rm var}} } } {\psi
                        _{{\mathop{\rm var}} } }}} \right. \kern-\nulldelimiterspace} {{\psi
                  _{{\mathop{\rm var}} } }}\right\rangle }}\right|},
  \end{equation}
  where
  \begin{equation}
    {\Delta _{i,\alpha }^{\dagger}=
    c_{i,\uparrow   }^{\dagger}c_{i+a _{\alpha },\downarrow}^{\dagger}-
    c_{i,\downarrow }^{\dagger}c_{i+a _{\alpha },\uparrow  }^{\dagger}}.
  \end{equation}
  The angular dependence of the real space correlations corresponds
  to those of the unprojected pairing symmetry. We have also
  checked that the value of $\Delta$ is independent of
  the choice of $\alpha$ and $\beta$.
\end{itemize}

\section{Results and discussion}
\label{sec:results}

\subsection{Half-filling}

We consider in this section the Heisenberg model (which is the limit
of the \tj model at half-filling, up to a constant) The comparison with
the large body of existing results for the Heisenberg model allows us to
benchmark the quality of our wave-function.

\begin{figure}
  \begin{center}
    \includegraphics[width=\figwidth]{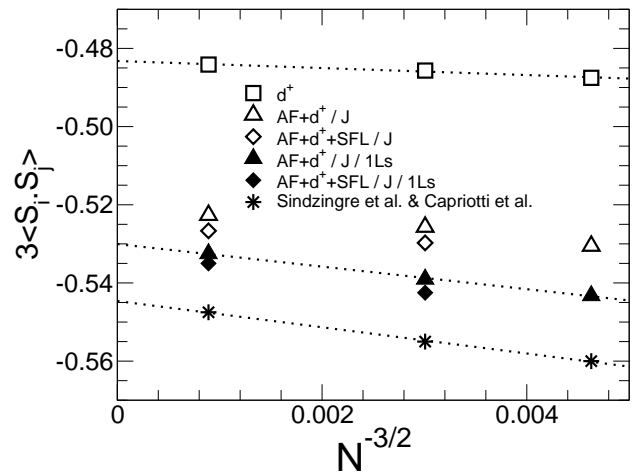}
    \caption{
      Energy per site $e=3 \langle \mathbf{S}_i\cdot\mathbf{S}_j\rangle$
      of the different variational wave functions for the Heisenberg model versus
      the system size $N^{-3/2}$, with $N=36,48,108$ sites.
      Ordered by increasing condensation energy we find:
      $d^{+}$ (open squares), $AF+d^{+}/J$ (open triangles),
      $AF+d^{+}+SFL/J$ (open diamonds), $AF+d^{+}/J/Ls$ (full triangles)
      and the $AF+d^{+}+SFL/J/Ls$ (full diamonds).
      The stars are the best estimates of the ground state energy available in the
      literature
      \cite{sindzingre_rvb_long_range_triangular,capriotti_spin_heisenberg_triangular}.
      \label{fig:scaling}}
  \end{center}
\end{figure}

Let us first briefly discuss the symmetry of the variational
parameters at half-filling. The variational magnetic field
minimizes the energy for the two degenerate $120^\circ$
configurations. We found that the BCS pairing symmetry in the
presence of $AF$ order is of $d^+$ type, whereas the $d^-$, the
$d_{x^2-y^2}$ and the $d_{xy}$ pairings have close but higher
energies. Finally, a staggered spin flux variational order
improves a little bit the energy. Interestingly, this latter
variational order is present in the ground state of the classical
Heisenberg model. However, this instability was only relevant at
half-filling, and the energy gain when $\delta>0$ is not
significant. The various energies for these wavefunctions are
shown in Fig.~\ref{fig:scaling}.
The $AF+d^{+}+SFL/J/Ls$ wave-function is thus the best
approximation, within our variational space, of the ground state
of the Heisenberg model. We compare its energy with other
estimates of the ground state energy in the literature (see
Fig.~\ref{fig:scaling} and Table \ref{tabel}).
\begin{table}
\caption{Comparison of the average energy $3\langle S_i \cdot S_j
 \rangle$ and the average magnetization $M_{AF}$ for the
 Heisenberg model (\tj model at half-filling)
 in different recent works for the $36$ site cluster
 and the extrapolation to the thermodynamic limit. The energy and the
 sublattice magnetization are measured
 for our best wave-function ($AF+d^++SFL/J/1Ls$) at half-filling.\label{tabel}}
\begin{tabular}{c c c}
  \hline
  \hline
    &  $\langle 3S_i \cdot S_j \rangle$ &  $M_{AF}$  \\
  \hline
   $36$ sites lattice  & \phantom{$6 \times 6$ sites lattice}  & \phantom{$6 \times 6$ sites lattice} \\
   our best wf          &  -0.543(1)                  &  0.38                       \\
   Capriotti et al.\cite{capriotti_spin_heisenberg_triangular}  &  -0.5581      &  0.406                      \\
   exact diag\cite{bernu_exact_diag_triangular,bernu_exact_diag_triangular_short}    &  -0.5604  &  0.400    \\
   \\
   $\infty \times \infty$ &                        &                              \\
   our best wf        &  -0.532(1)               &  0.36                        \\
   spin-wave results\cite{capriotti_spin_heisenberg_triangular}    &  -0.540                  &  0.25        \\
   Capriotti et al.\cite{capriotti_spin_heisenberg_triangular} &  -0.545       &  0.21                        \\
  \hline
  \hline
\end{tabular}
\end{table}
The mixture of $AF$ and $d^{+}$ instabilities is improving a lot
the energy, and our wave-function has significantly lower energies
than the simple $d^+$ wave-function, and has energies very close
to the best ones available. More precisely, we find in the
thermodynamic limit an energy per site of $e=-0.52J$ for our
variational wave-function. Applying one Lanczos step leads to a
small further improvement of the energy to $e=-0.53J$. A summary
of the energies and of the $120^\circ$ magnetization for the
Heisenberg model are given in Table~\ref{tabel} and in
Fig.~\ref{fig:scaling}.

Inspection of these results shows that our wave-function has in
the thermodynamic limit an energy only $0.013J$ higher than the
estimates of more sophisticated, but restricted to the undoped
case, methods\cite{capriotti_spin_heisenberg_triangular}. Indeed,
these latter methods use pure spin variational wave-function as a
starting point, restoring quantum fluctuations with a
non-variational method. Let us point out however that these
methods are not giving an upper bound on the true ground state
energy, so the ground state energy could in principle lie between
this result and our variational one. Since our variational
wavefunction already gives an excellent energy it would be
interesting to check how the (non variational) methods used to
improve the energy starting with a much cruder variational
starting point would work with our variational wavefunction and
which energy it would give. We leave this point for future
investigation however.

Our wavefunction shows a reduced but finite magnetic order that
survives in the triangular Heisenberg  antiferromagnet (THA). The
$120^\circ$ magnetization of our wavefunction is reduced by the
BCS pairing down to $72\%$ of the classical value (see
Fig.~\ref{fig:M}) which is somewhat larger than the spin-wave
result. Thus in addition to having an excellent energy, our
wavefunction seems to capture the physics of the ground state of
the Heisenberg system correctly. Let us note that the BCS order of
the wave-function is destroyed by the full Gutzwiller projector at
half-filling. So, despite the presence of a variational
superconducting order parameter, the system is of course not
superconducting at half filling. Somehow the BCS variational
parameter helps to form singlets, which reduces the amplitude of
the AF order. This is very similar to what happens for the \tj
model on the square lattice: the inclusion of a superconducting
gap decreases the energy and decreases also the magnetization from
$M \approx 0.9$ down to $M \approx 0.7$, which is somewhat larger
than the best QMC estimates ($M\approx 0.6$, see
Refs.~\cite{giamarchi_tjhub,Sandvik_2d_HB}). Thus the
wave-function mixing magnetism and a RVB gap seem to be
interesting variationally, both in the square and triangular
lattice, to restore spin fluctuations that were frozen in the pure
classical magnetic wave-function. For the triangular lattice, the
present work is the first attempt to reproduce the magnetic order
in the THA in terms of a fermionic representation, which gives
results in good agreement with other methods. The great advantage
of this approach is of course that the fermionic language allows
to directly consider the case of hole and electron doping in the
AF background, which is the case we consider in the following
sections.

\subsection{Electron doping: $n \in [1,2]$}

Very few results exist away from half-filling, so in order to have
a point of comparison for our variational approach we will compare
it with exact diagonalizations on very small clusters. Having
ascertained that our wavefunction is indeed in good agreement with
the exact results on a small cluster, we can then use it with
confidence to describe much larger systems and extract the physics
of the thermodynamic limit.

\begin{figure}
  \begin{center}
    \includegraphics[width=\figwidth]{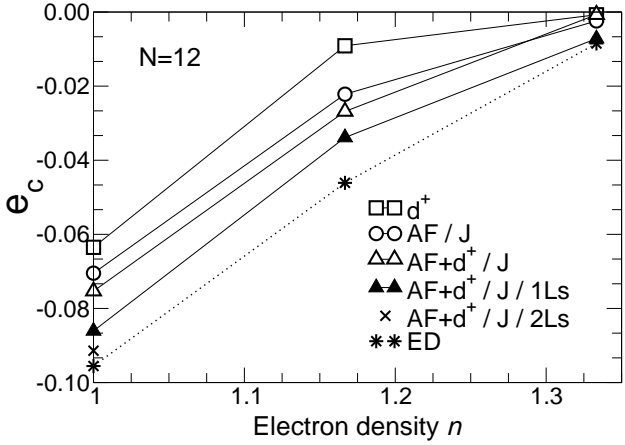}
    \caption{Condensation energy per site versus the electron doping for a
      $12$ site cluster for the different variational wave-functions.
      We have done exact diagonalization (ED) for a 12 sites cluster
      with same periodic boundary conditions.
      \label{fig:12sites}}
  \end{center}
\end{figure}

Therefore, we start by comparing on a $12$ site cluster different
wave-functions with the exact-diagonalization results for the case
of electron doping (see Fig.~\ref{fig:12sites}), since larger
lattices are not readily available. Interestingly, it was found
that even with only $2$ Lanczos steps on our best wave-function
($AF+d^{+}/J$) the energy has almost converged to the exact ground
state energy at half-filling.

Note that small system size is the worst possible case for a VMC
method since the simple variational wavefunction is not expected
to reproduce well the short distance correlations, as we fix the
long-range magnetic correlations in our variational ansatz by
imposing an on-site magnetic field, but we do not introduce short
range corrections. Variational Monte-Carlo instead focus on the
long distance properties, which will become dominant in the energy
as the lattice size increases. Nevertheless, the short range
correlations contributes significantly to the energy on small
lattices. One can thus expect on general grounds the energies to
become increasingly good as the system size increases, provided
that the correct long range order has been implemented in the
wavefunction. The Lanczos iterations allow to correct this local
structure of the wave-function. Here we see that by changing this
local structure our wave-function is converging very fast to the
ground state. This is a good indication that even away from half
filling our wavefunction is quite efficient in capturing the
physics of the system. Actually, the variance of the energy per
site $\sigma^2$ reaches its maximum value for doping
$x=\frac{1}{3}$ ($\sigma^2=0.006$), but applying one Lanczos step
reduces drastically the variance : $\sigma^2=0.0004$. At
half-filling, a variance-energy plot for the three functions
$AF+d^{+}/J/pLs$ ($p=0,1,2$) allows to extrapolate the energy at
$0$ variance, and we get an energy per site of $e=-0.61(1)$, which
is very close to the exact result $e=-0.6103$.

\begin{figure}
  \begin{center}
    \includegraphics[width=\figwidth]{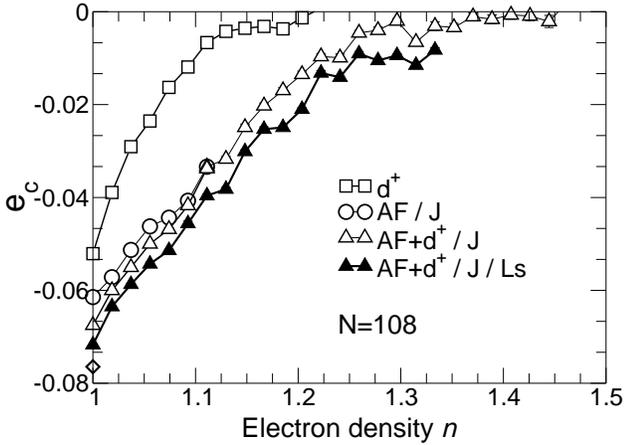}
    \caption{Condensation energy per site $e_c$ versus electron doping for
      the $108$ sites lattice. We show different wave-functions and also the best
      estimate in
      the literature\cite{capriotti_spin_heisenberg_triangular} at
      half-filling (open diamond).
      \label{fig:Ec}}
  \end{center}
\end{figure}

\begin{figure}
  \begin{center}
    \includegraphics[width=\figwidth]{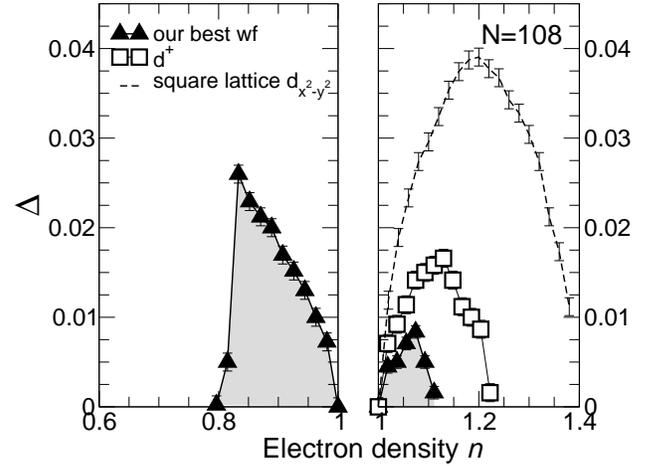}
    \caption{Superconducting order parameter $\Delta $ for
      a $108$ site triangular cluster in our best wave-function (full triangles),
      in the $d^+$ wave-function (open squares).
      For comparison we show the amplitude of the d-wave gap
      in a $10\times 10$ {\em square} lattice (dashed line).}
    \label{fig:delta}
  \end{center}
\end{figure}

Let us now use our wavefunction to describe large systems away
from half filling. We now focus on a $108$ site cluster, which is
the largest cluster we can treat with a reasonable effort. We have
first measured the condensation energy per site (see
Fig.~\ref{fig:Ec}) for different types of instabilities.
Very interestingly, the $AF/J$ is even better than a simple $\DID$
RVB state. Moreover, the RVB order is only weakly increasing the
condensation energy in presence of the antiferromagnetic
background ($AF+d^+/J$). This is suggesting that superconductivity
is only weakly present in the \tj model when $n>1$ which is also
confirmed by the measurement of the superconducting gap (see
Fig.~\ref{fig:delta}).
The superconducting order of our best wave-function is
approximately $4$ times smaller in amplitude and in range of
stability than the d-wave pairing in a $10 \times 10$ {\em square}
lattice with the same boundary conditions. For electron doping
$\delta>0.04$ we find that the $d^+$ BCS pairing symmetry has the
same energies as the $d^-$ one, and also as the wavefunctions with
$d_{x^2-y^2}$ and $d_{xy}$ pairings.

\begin{figure}
  \begin{center}
    \includegraphics[width=\figwidth]{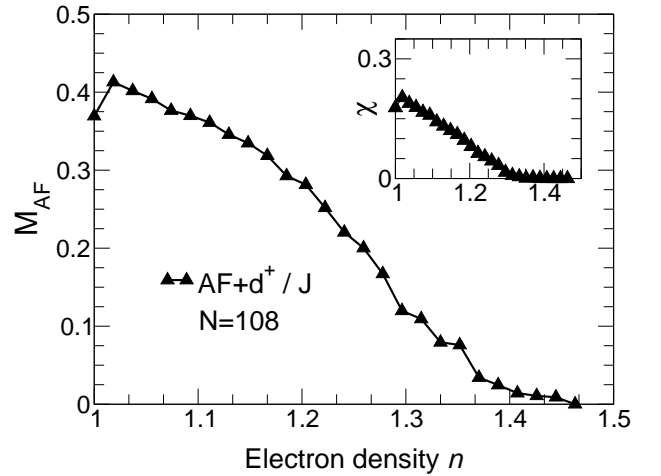}
    \caption{Amplitude of the $120^{\circ }$ magnetic order $M_{AF}$
      measured in the $AF+d^+/J$ for a $108$ site cluster. Inset: the
      amplitude of the staggered spin-current $\chi$ in the same wave-function.
      \label{fig:M}}
  \end{center}
\end{figure}

Very strikingly, the $120^\circ$ magnetic order parameter is
surviving up to high doping $\delta=0.4$, see Fig.~\ref{fig:M}.
Long-range magnetic order at finite doping is potentially caused
by a limitation of the VMC method, in that it is not possible in
our calculation to model wavefunction with a finite correlation
length, i.e. short range $120^\circ$ magnetic order. In our
calculation, we can either totally suppress the long-range
$120^\circ$ and get back to the Gutzwiller wave-function, or use
the long-range $120^\circ$ magnetic order that is highly
stabilized by the potential energy. No intermediate scenario, such
as incommensurate structures, is yet available in our
calculations, but one can only expect the optimization of the
magnetic structure to increase the region of stability for
magnetism. We interpret this finding as an indication that the
hole motion is not drastically modified by the presence of the
non-collinear magnetic structure, so that short range magnetic
correlations will survive up to high electron doping.
For the $t{-}J^z$ model on the square lattice, it is commonly
understood that the Ising N\'eel order is not surviving high
doping because of its costs in kinetic energy: whenever a hole
wants to move in a antiferromagnetic spin background, it generates
a ferromagnetic cloud. Therefore, good kinetic energies and N\'eel
Ising order are not compatible. In our case, the $3$-sublattice
order imposes no such constraint on the kinetic energy of the
holes, because of the $120^\circ$ structure. We see this fact in
our wave-function energies: the potential energy is improved when
starting from the Gutzwiller wavefunction and adding $120^\circ$
correlations, but the kinetic energy is unchanged. Our best
wavefunction has a better potential energy than the Gutzwiller
wavefunction (and also than the different CFP wavefunction), but
it also keeps the same kinetic energy. Therefore, this
qualitatively explains why one can stabilize the $3$-sublattice
magnetic phase for a large set of $J$ values. Finally, we note
that the staggered spin-current pattern is also present for doping
$\delta=\left[0,0.3\right]$ (see Fig.~\ref{fig:M}).

Interestingly, the Jastrow variational parameter $\alpha$
(\ref{eq:jastrow}) is changing sign at $\delta=0.4$ when $M_{AF}
\approx 0$ : for $\delta<0.4$ ($\delta>0.4$) the Jastrow factor
favors classical Ising (ferro-) anti-ferromagnetic states. The VMC
results show that the competition between the classical Ising
configuration on the triangular lattice and the $120^\circ$ order
is improving the energy. We argue that the classical Jastrow
simulates with a good approximation quantum fluctuations around
the $120^\circ$ order. Note that the Jastrow does not play the
same role as the BCS pairing: the BCS pairing forms configurations
of resonating singlets, and the Jastrow factor forms classical
Ising configurations. It is also worth noting that at higher
doping the Jastrow parameter is leading to a small condensation
energy of $0.01t$ for a large range of doping
($\delta=\left[0.4,0.8\right]$). It was checked that this gain in
energy does not decrease with the size of the lattice and is also
present for a square lattice geometry. We found also that for the
small clusters ($12$ and $48$ sites) the system was gaining a
significant amount of energy when having a weak ferromagnetic
polarization. Therefore, this is suggesting that the Gutzwiller
wavefunction is not the best approximation of the ground state of
the \tj model in the high doping limit. Nevertheless, the Jastrow
factor does not introduce long-range correlation and the
variational wavefunction we introduce here is still a
Fermi-Liquid.

Note that the variance of the energy per site $\sigma^2$ reaches
its maximum value for doping $\delta=0.4$ for the $AF+d^+/J$ with
$\sigma^2=0.0008$, and applying one Lanczos step leads to
$\sigma^2=0.0004$.

\subsection{Hole doping: $n \in [0,1]$}

\begin{figure}
  \begin{center}
    \includegraphics[width=\figwidth]{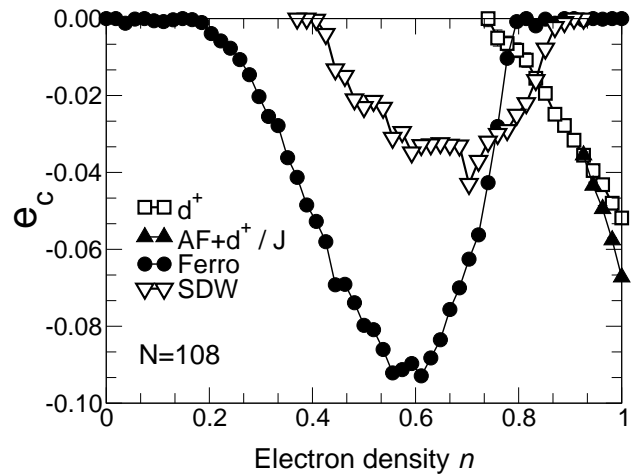}
    \caption{Condensation energy per site $e_{c}$ for different wavefunctions in
      a $108$ site cluster.\label{fig:Ecm}}
  \end{center}
\end{figure}

\begin{figure}
  \begin{center}
    \includegraphics[width=\figwidth]{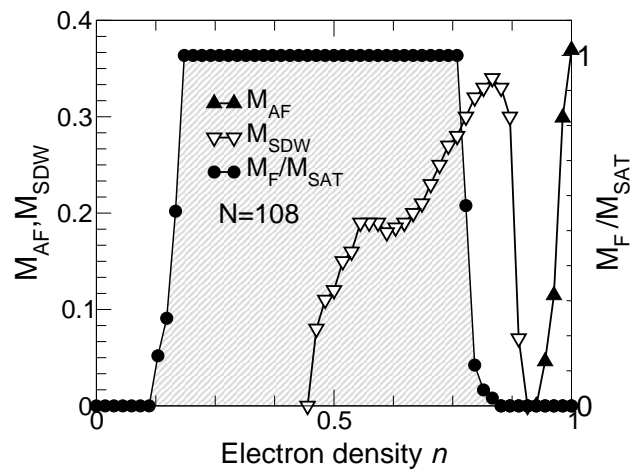}
    \caption{Amplitude of the $120^{\circ }$ magnetic order measured
      in our best wavefunction for the $108$ site cluster (left scale, full
      triangles) and the ratio of the polarization $M_F$ on the saturated
      polarization $M_{sat}$ in our best wavefunction (right scale, full circles).
      We show also the absolute magnetization $M_{SDW}$ for the spin density
      wave wavefunction (left scale, see also Fig.~\ref{fig:spinz}).
      \label{fig:Mm}}
  \end{center}
\end{figure}

For hole doping the scenario is strikingly different. The
$120^\circ$ order is weakened in a strong $d^+$ RVB background and
disappears at doping $\delta=0.08$ (see Fig.~\ref{fig:Ecm} and
Fig.~\ref{fig:Mm}).
\begin{figure}
  \begin{center}
    \includegraphics[width=\figwidth]{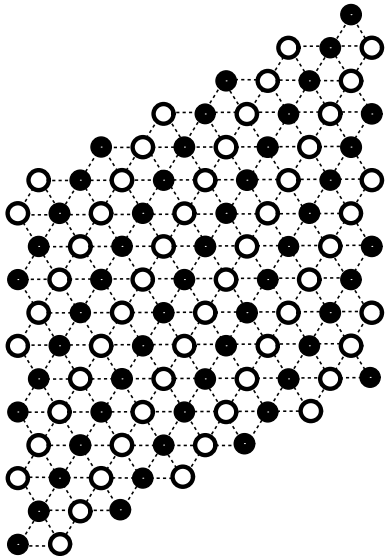}
    \caption{On-site magnetization for each site of a $108$ site lattice
    for the spin density wave wavefunction.
    Open (filled) circles denotes down (up) spins.
    The size of each circle is proportional to the
    respective amplitude of the on-site magnetization.
    We find that the spins forms a stripe-like pattern,
    alternating ferromagnetic bonds in the $\mathbf{a_2}$
    direction, and antiferromagnetic bond in the two other
    directions. The average on the lattice sites
    of the absolute value of the local magnetization
    is shown in Fig.~\ref{fig:Mm}.}
  \label{fig:spinz}
  \end{center}
\end{figure}
When superconductivity disappears, there is a first order
transition to a commensurate spin density wave. No coexistence
between superconductivity and the spin density wave was found.
Then, a ferromagnetic phases emerges with a strong gain of
condensation energy. Indeed the polarized states are leading to a
strong gain in kinetic energy. This can be understood in the
simple picture of the Stoner model, which gives a critical onsite
repulsion related to the density of states:
$U_{cr}^F=1/\rho(\epsilon_F)$. Ferromagnetism becomes favorable if
$\epsilon_F$ is sitting at a sharp peak of $\rho(\epsilon)$. In
the triangular lattice the tight binding (TB) density of states is
strongly asymmetric and has a sharp peak at the $n=0.5$ electronic
density lying at the Van Hove singularity. Note also that the
simple \tj model of a $3$ site cluster with $2$ electrons shows
that in the $t>0$ the ground state is a singlet, whereas the
ground state is a triplet in the $t<0$ case. This shows that the
negative sign of $t$ with hole doping is inducing ferromagnetic
correlations on a very small cluster. We find again trace of these
correlations and ferromagnetic tendencies in the range of
electronic density $n\in[0.2,0.8]$ in our $108$ site lattice. Such
a ferromagnetic instability was also predicted in
Ref.~\onlinecite{watanabe_VMC_TJ_triangular} by comparing the
energy of the RVB wavefunction with an analytical calculation of
the energy of the fully polarized state. We see that minimizing
the energy by changing the variational onsite magnetic field leads
to similar results.

Moreover, at $\delta=0.5$ doping, there is a nesting of the Fermi
surface, with three possible $\mathbf{Q}$ vectors. Thus, it is
reasonable to expect that a particle-hole instability of
corresponding pitch vector $\mathbf{Q}$ is stabilized close to
this doping. We have investigated the following instabilities : a
commensurate charge density wave, and a spin density wave.
Interestingly, the commensurate spin density wave was stabilized.
Indeed, we have found that the scattering between the $\mathbf{k}$
and $\mathbf{k+Q}$ vectors introduced in the Hamiltonian $H_{SDW}$
(particle-hole channel) allow to gain kinetic energy in the range
of doping $\delta=[0.15,0.6]$. For sake of simplicity, we have
only considered the mean-field Hamiltonian containing one of the
three possible nesting vectors : $\QQ$. Finally, the phase is
stabilized, when compared to the RVB and ferromagnetic phases, in
the window $\delta=[0.16,0.24]$. Nonetheless, no coexistence
between superconductivity and the spin density wave was found :
the energy is minimized either for $\left(\Delta_k=0,f(Q)\neq
0\right)$, or $\left(\Delta_k\neq 0,f(Q)=0\right)$ depending on
the doping, with $\Delta_k$ of $\DID$ symmetry type in the latter
case. Measuring the on-site magnetization value, we found that the
spin density wave is forming a collinear stripe-like pattern in
the spins degrees of freedom, whereas the charge is found to be
uniformly distributed among the lattice sites, as expected (see
Fig.~\ref{fig:spinz}). The amplitude of the on-site magnetization
is shown in Fig.~\ref{fig:Mm} as a function of doping.

\subsection{Phase diagram of the model}

\begin{figure}
  \begin{center}
    \includegraphics[width=\figwidth]{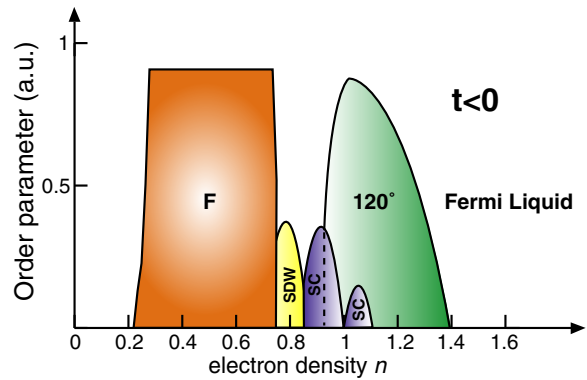}
    \caption{(color online) Cartoon picture of the phase diagram
    of the \tj model we get with $t<0$. Here
      we sketch on an arbitrary scale
      the order parameter amplitude of the $120^\circ$
      magnetic phase, the ferromagnetic phase ($F$), the
      superconducting $\DID$ phase ($SC$), the commensurate
      spin density wave ($SDW$). Note
      that for electron density $n>1.04$ and $n<0.96$,
      the energy is degenerated, within the error bars due to the Monte-Carlo sampling,
      with the pairings $d_{x^2-y^2}$, $d_{xy}$ and
      $d_{x^{2}-y^{2}}-id_{xy}$. The pitch vector of the commensurate spin density
      wave is $\QQ$, and this phase is depicted more in details in Fig.~\ref{fig:spinz}.
      } \label{fig:diag}
  \end{center}
\end{figure}

Based on our wavefunction we can now give the phase diagram for
the doped system on the triangular lattice. The phase diagram,
summarizing the various instabilities discussed in the previous
sections is show in Fig.~\ref{fig:diag}.
This phase diagram prompts for several comments. First one notices
immediately that the competition between magnetism and
superconductivity in this model depends crucially on the sign of
the hopping integral (see Fig.~\ref{fig:diag}).

For both hole and electron doping, the triangular lattice has a
very different phase diagram from the square lattice one
\footnote{ In the square lattice the sign of $t$ plays no role
because of the particle-hole symmetry of a bipartite lattice.}. In
the square lattice, the AF order disappears at $\delta=0.1$ and
the d-wave RVB dies at $\delta=0.4$ for the same value of $J$. In
the triangular lattice, a similar stability of superconductivity
exists on the hole side, but the electronically doped side is
resolutely dominated by antiferromagnetic instabilities. Our
results, based on an improved class of wavefunctions, present
marked differences with previous approximate results for the doped
system. On the electron side, mean-field theories would have
suggested that the long-range magnetic order state undergoes a
first order phase transition\cite{wang_mf_triangular} into a
uniform $d^+$ superconducting state at $\delta\approx3\%$ for
values of $J$ similar to those considered here. A rationalization
of these results would be that the frustration of the lattice,
which was from the start the motivation of RVB as a competing
state, disfavors magnetic order. Our results, where the Gutzwiller
projection is treated exactly within the residual error bars due
to the statistics, are in strong disagreement with this mean-field
theory. Contrarily to the mean-field result, magnetism is dominant
and the superconducting order is not favored on the electronic
side. In addition, the \tj model on the triangular lattice was
expected to have a strong and large RVB instability, since the
coordination number is higher than on most of the other lattices,
and naively we would expect this to provide an easy way to form
singlets. In this work we show that it is not the case: for
electron doping the system is magnetic, and for hole doping the
system is superconducting, but the range of superconductivity is
not extraordinary large ($\delta<0.16$), and smaller than on the
square lattice.

Previous variational approaches\cite{watanabe_VMC_TJ_triangular}
were restricted to pure superconducting wavefunctions $\DID$ on a
$t-t'$ square lattice with $t=t'$. In that work it was found that
superconductivity is stabilized up to electron doping $\delta
\approx 0.24$ and hole doping $\delta \approx 0.2$ for similar
albeit slightly different values of $J/t$ ($J/t=0.3$). In our
work, for the case of electron doping, which corresponds to the
doping in the cobaltite experiments, our phase diagram, using the
larger class of wavefunctions, is clearly completely different
from this previous result, and the stabilization of the
superconductivity in that case was clearly an artefact of the too
restricted variational subspace. As show in Fig.~\ref{fig:diag}
superconductivity is strongly weakened by the presence of
3-sublattice magnetization and is present only in the range
of electron doping $\delta=[0,0.12]$. On the contrary, for the case of hole doping,
superconductivity had higher energy than ferromagnetic and spin
density wave phases for $\delta>0.16$. We thus confirm that the
previous results are not an artefact of their restricted
variational subspace, and find an acceptable agreement for the
phase diagram. However we emphasize the presence of the spin
density wave wavefunction that was not considered in the mentioned
work and implies a small reduction of the superconductivity range.

Our calculation thus clearly prompts for a reexaminations of the
arguments on the nature of superconductivity in a frustrated
lattice. Clearly the non-collinear nature of the order parameter
helps making the AF order much more stable to electron doping than
initially anticipated. Understanding such issues is of course a
very crucial and challenging question. Moreover, on the triangular
lattice, no significant enhanced cooperative effect between
magnetism and superconductivity seems to be observed: the electron
doped side has a magnetic signature, and the hole doped side a
superconducting one, but the two orders seem to exclude each other
as much as they can, contrarily to what happens for the square
lattice. Even in the parts of the phase diagram where coexistence
is observed, coexistence between magnetism and superconductivity
in the electron doped case shows again that superconductivity is
decreased in the presence of strong long range magnetic
correlations.

\section{Conclusion} \label{sec:conclusion}

In this paper we have presented a variational Monte-Carlo study of
the \tj ($J/t=0.4$ and $t<0$) model on the triangular lattice,
using extended wavefunction containing both superconductivity and
non-collinear magnetism, as well as flux phase instabilities. The
method we used to construct and sample the wavefunction is quite
general and applicable to other lattices (honeycomb, kagome,
ladders...) as well as other symmetries (e.g. triplet
superconductivity). It thus provides a general framework to tackle
the competition between antiferromagnetism and superconductivity
in frustrated systems. We obtained very good variational energies
at half-filling when comparing with other more sophisticated
methods, specialized to the half-filled case. The fermionic
representation of our wavefunction allows to consider hole and
electron doping. The most stable pairing corresponds to singlet
pairing. We find that $\DID$ superconductivity is only weakly
stabilized for electron doping in a very small window
($\delta=[0,0.12]$) and is much stronger and also appears in a
wider range ($\delta=[0,0.16]$) in the case of hole doping. A
commensurate spin density wave phase is leading to a gain in
kinetic energy and is stabilized in the small window
$\delta=[0.16,0.24]$ hole doping. Finally, ferromagnetism emerges in a wide
range for hole doping $\delta=[0.24,0.8]$. Very surprisingly, the
3-sublattice magnetism which is present at half-filling extends to
a very wide range of electron doping $\delta=[0,0.4]$ and is
suppressed very fast in the case of hole doping $\delta=[0,0.08]$.
The large extent of $120^\circ$ order for electron doping is
responsible for the suppression of superconductivity. This feature
was neither observed in previous VMC calculations, nor predicted
by the mean field theories, and prompts for a reexamination of the
question of the stability of magnetic order on a triangular
system.

Our results show that, for electron doping, the square and
triangular lattices behave in a very different way. For the square
lattice, the \tj Hamiltonian finds a domain of stability of
superconductivity and a pairing symmetry that is very consistent
with other methods. It is thus a natural candidate to investigate
superconducting phases in systems like the cuprates. For the case
of the triangular lattice, the predicted phase diagram is
dominated by antiferromagnetic instabilities, and
superconductivity, albeit slightly present, is strongly
suppressed. This clearly indicates that, contrarily to what was
suggested by mean-field and previous variational calculations, the
\tj model itself is not a good starting point to tackle the
superconductivity of the cobaltite compounds, where
superconductivity is observed in the range of electron density
$n=\left[1+\frac{1}{4},1+\frac{1}{3} \right]$. This model must be
completed by additional ingredients to obtain a faithful
description of the experimental system. Two missing ingredients in
the simple \tj model could solve this discrepancy and perhaps
allow to obtain a superconducting instability. On one hand, a
strong Coulomb repulsion is expected in this type of compound.
Such a long-range interaction is not taken into account in the
\tj model. Thus a coulomb $V$ term should be added to get a
$t{-}J{-}V$ model. On the other hand, in this paper, we have used a
single-band model as a first step to study the Co-based oxides.
However, it is quite possible that the multi-band effect plays an
essential role for superconductivity\cite{koshibae_multiband_triangular}.
The interaction between the three bands of the compound could play a
non trivial role in the physics of the \tj model. Therefore, a
study of the 3-band model could also be of interest. Such
an analysis can be done by extending the methods exposed in this
paper to these more complicated models.

\acknowledgments We are grateful to Federico Becca and Antoine
Georges for useful discussions. This work was supported by the
Swiss National Fund and by NCCR MaNEP.

\end{document}